\documentclass[usenatbib]{mn2e}
\usepackage{multirow}
\usepackage{rotating}
\usepackage{colortbl}
\usepackage{color}
\usepackage[fleqn]{amsmath}
\usepackage{amssymb}
\usepackage{amsfonts}
\usepackage{verbatim}
\usepackage{scalefnt}
\usepackage[percent]{overpic}

\newcommand{\araa}{ARA\&A}
\newcommand{\apj}{ApJ}
\newcommand{\apjl}{ApJ}
\newcommand{\aap}{A\&A}
\newcommand{\mnras}{MNRAS}

\newcommand{\sfrn}{{\rm SFR}_{100{\rm pc}}}

\newcommand{\sfrg}{{\rm SFR}_{5{\rm kpc}}}
\newcommand{\beq}{\begin{equation}}
\newcommand{\eeq}{\end{equation}}
\newcommand{\ba}{\begin{eqnarray}}
\newcommand{\ea}{\end{eqnarray}}
\def\spose#1{\hbox to 0pt{#1\hss}}
\newcommand{\lta}{\mathrel{\spose{\lower 3pt\hbox{$\mathchar'218$}}
      \raise 2.0pt\hbox{$\mathchar"13C$}}}
\newcommand{\gta}{\mathrel{\spose{\lower 3pt\hbox{$\mathchar"218$}}
      \raise 2.0pt\hbox{$\mathchar"13E$}}}

\newcommand{\comments}[1]{} 

\title[BH accretion and star formation rate]{Black hole accretion versus star formation rate: theory confronts observations}

\author[Volonteri et al.]{Marta Volonteri,$^{1}$\thanks{E-mail: martav@iap.fr}  Pedro R. Capelo$^{2}$, Hagai Netzer$^{3}$,\newauthor Jillian Bellovary$^{4}$, Massimo Dotti$^{5}$, Fabio Governato$^6$\\
$^1$Sorbonne Universites, UPMC Univ Paris 06 et CNRS, UMR 7095, Institut dÕAstrophysique de Paris, F-75014, Paris, France\\
$^2$Department of Astronomy, University of Michigan, Ann Arbor, MI 48109, USA\\
$^3$School of Physics and Astronomy, The Sackler Faculty of Exact Sciences, Tel-Aviv University, Tel-Aviv 69978, Israel\\
$^4$Department of Physics and Astronomy, Vanderbilt University, Nashville, TN 37235, USA\\
$^5$Dipartimento di Fisica G. Occhialini, Universit$\grave{a}$ degli Studi di Milano Bicocca, Piazza della Scienza 3, I-20126 Milano, Italy\\
$^6$Department of Astronomy, University of Washington, Box 351580, Seattle, WA 98195, USA}

\begin{document}

\maketitle

\begin{abstract}
We use a suite of hydrodynamical simulations of galaxy mergers to  compare star formation rate (SFR) and black hole accretion rate (BHAR) for galaxies before the interaction 
(`stochastic' phase), during the `merger' proper,  lasting $\sim 0.2-0.3$~Gyr, and in the `remnant' phase. We calculate the bi-variate distribution of SFR and BHAR and define the regions in the SFR-BHAR plane that the three phases occupy. No strong correlation between  BHAR and galaxy-wide SFR is found. A possible exception are galaxies with the highest SFR and the highest  BHAR.  We also bin the data in the same way used in several observational studies, by either measuring the mean SFR for AGN in different luminosity bins, or the mean BHAR for galaxies in bins of SFR.  We find that the apparent contradiction or SFR versus BHAR for observed samples of AGN and star forming galaxies is actually caused by binning effects.  The two types of samples use different projections of the full bi-variate distribution, and the full information would lead to unambiguous interpretation. We also find that a galaxy can be classified as AGN-dominated up to 1.5~Gyr after the merger-driven starburst  took place.  Our study is consistent with the suggestion that most low-luminosity AGN hosts do not show morphological disturbances.
\end{abstract}


\section{Introduction}\label{sec:Introduction}

The correlation (or lack thereof) between the black hole accretion rate (BHAR) and the star formation rate (SFR) of their host galaxies has been the subject of numerous
 investigations  \citep[e.g.,][and references therein]{2009MNRAS.399.1907N,Mullaney2012,2012A&A...545A..45R,2012ApJ...760L..15H,2013A&A...560A..72R,2014ApJ...791...34N,2014ApJ...782....9H}. One of the drivers behind these studies is the empirical 
correlation between BH mass and bulge mass in the local universe \citep[e.g.,][]{MarconiHunt2003,Haring2004,Gultekin2009} which suggests that the BH 
and stellar bulge have assembled in tandem (co-evolution). In the strictest view of co-evolution, to obtain a BH mass proportional to the bulge mass, BHAR should 
be proportional to SFR (or, at least, the SFR that builds-up the bulge). Indeed, the cosmic total SFR and BHAR (i.e., the rate per unit comoving volume) seem to track each other 
at least to $z\sim 3$ \citep{2004ApJ...613..109H,Merlonietal2004,2008ApJ...679..118S,2009ApJ...696..396S,2014ARA&A..52..415M}. 

When  SFR and  BHAR are compared source by source, the connection appears to be weak, unless only the SFR in the central region is taken into account 
\citep[$<$1~kpc,][and references therein]{2012ApJ...746..168D}. A comprehensive work of this type by \cite{2012A&A...545A..45R}, shows that at low AGN luminosities,
 BHAR and SFR are uncorrelated, while at high AGN luminosities a significant correlation emerges. On the other  hand, Mullaney et al. (2012) and Chen et al. (2013) study a sample of mass-selected and star-forming galaxies respectively and suggest that the average BHAR correlates  well with the average SFR, once the shorter variability time scales of BHAR with respect to SFR are taken  into account \citep[see also][]{2014ApJ...782....9H}.

In a companion paper (Volonteri et al. 2015) we have used new simulations of galaxy mergers and investigated  the temporal correlation between SFR and BHAR, and their respective variability. We found that BHAR and nuclear ($<100$ pc) SFR are well correlated and vary on similar timescales. However, we found that galaxy-wide ($<5$ kpc) SFR, which is used in statistical studies \citep{2012A&A...545A..45R,Mullaney2012,2013ApJ...773....3C,2015arXiv150107602D}, and BHAR are typically temporally uncorrelated, and have different variability timescales, except during the short-lived merger proper ($\sim 0.2-0.3$~Gyr). We here compare SFR and BHAR by referring to the luminosity resulting from these processes ($L_{\rm SFR}$ and $L_{\rm AGN}$ respectively), and aim at providing a framework for comparison with observations. 

In this Letter we show how very detailed numerical simulation of galaxy mergers confirm the recent observational finding, provided the simulation results are treated in the same
way as observations. In particular we show that accounting for the different durations of the various stages of the mergers, and binning
the results in a way similar to the observational methods, can bring the results very close to the observed correlations and reconciles seemingly conflicting  results. The results presented here are based on our previous detailed work published in \cite{2015MNRAS.447.2123C} and Volonteri et al. (2015). We refer the reader to those papers for a complete description of the simulations, of the physical implementation and of the analysis.

\section{Comparison with observations}\label{sec:obscomp}

\begin{figure*}
\centering
\includegraphics[width=\columnwidth,angle=0]{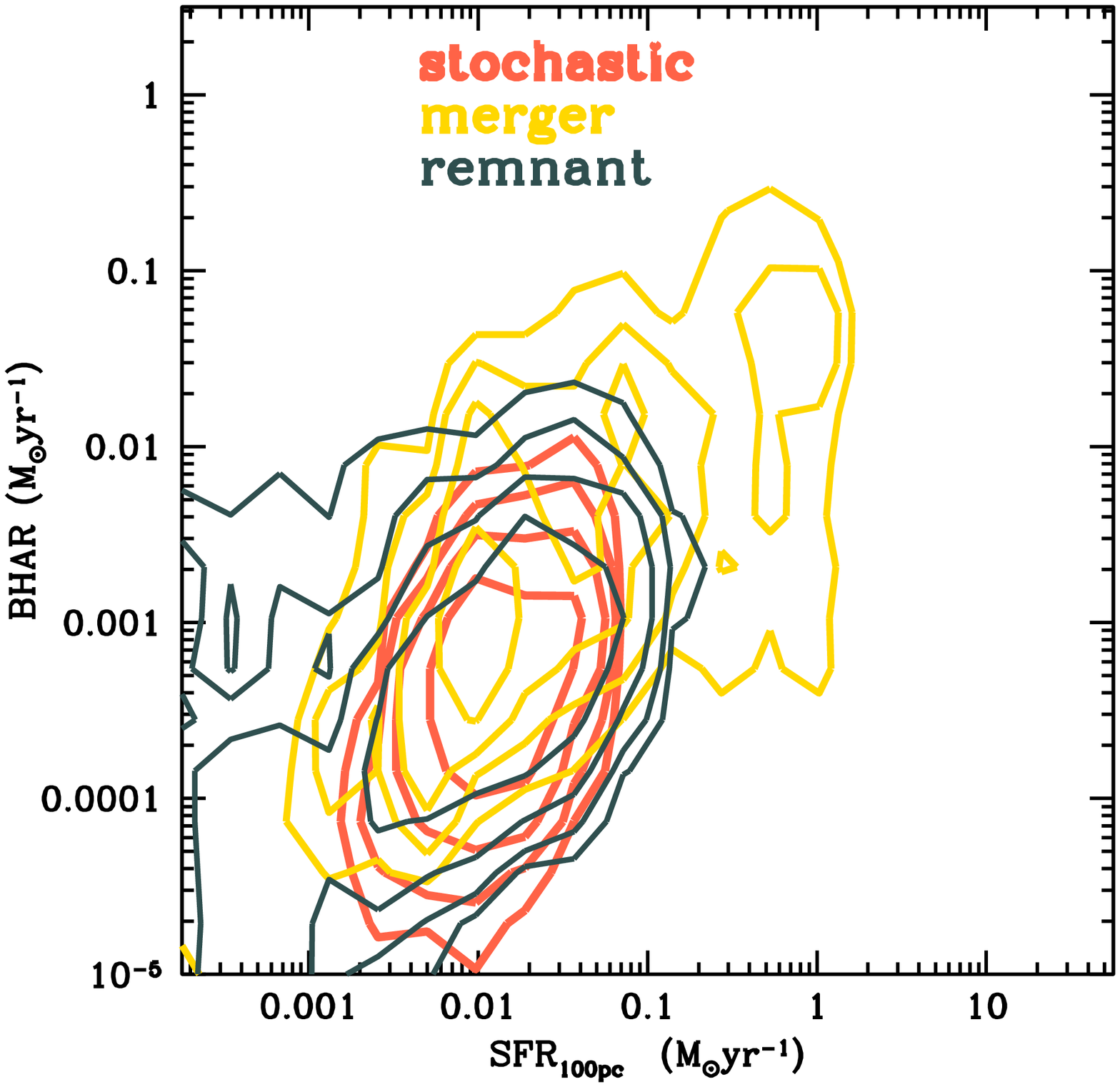}
\includegraphics[width=\columnwidth,angle=0]{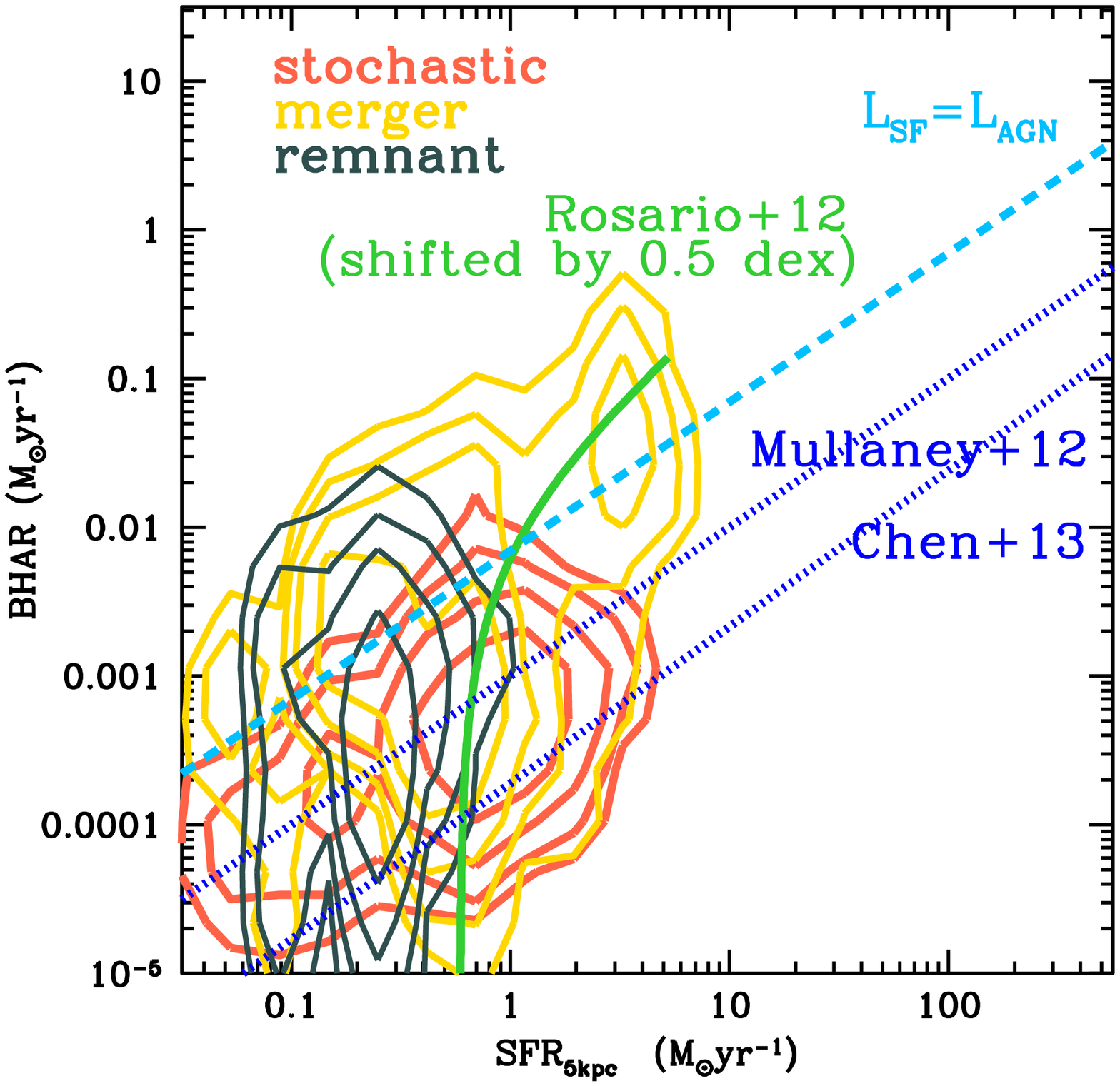}
\caption{BHAR versus SFR within 100~pc (left), and within 5~kpc (right). Contours are based on ten equally-spaced logarithmic levels, dropping the six lowest levels for clarity.  The SFR is an average over the 100 Myr before the time-step used for the BHAR calculation. We distinguish stochastic (red), merger  (gold) and remnant (dark gray) phases.   We show curves from Rosario et al. (2012, AGN, green, see section 3.2 for details), Mullaney et al. 2012  (mass-selected galaxies, upper blue dotted line) and Chen et al. (2013, star-forming galaxies, lower blue dotted line).  We also mark the line separating AGN and SF dominated regions (light blue dashed line). }
\label{fig:sfr_bhar_2w_app}
\end{figure*}

The comparison of the numerical simulations with observed correlations has some caveats.   As discussed by \cite{2014MNRAS.437.3373N}, measurements of SFRs are based on indicators that last $\gtrsim$100~Myr, while the AGN  luminosity, hence the BHAR, is measured instantaneously. Because of this, all our results concerning SFR represent averages over 100 Myr.  The SFR is converted to SF luminosity, $L_{\rm SFR}$, by assuming that one solar mass per year corresponds to $10^{10}$ solar luminosities. BHAR is averaged instead over 1~Myr (see Volonteri et al. 2015 for a case where BHAR and SFR are measured over the same time interval). We distinguish between SF dominated galaxies ($L_{SF} > L_{AGN}$) and AGN dominated galaxies ($L_{SF} < L_{AGN}$). 

Dense gas in the host galaxy may cause some obscuration, especially during the merger phase. We estimated the column density for a 1:2 merger, in cylinders of radius $r = 100$~pc centred on the BHs, and found that it varies between $\sim10^{22}$ (face-on) and $\sim10^{23}$~cm$^{-2}$ (edge-on), with an increase by a factor of 2 during the merger proper.  Using the WebPIMMS interface (with a photon index $\Gamma = 1.4$), the average (maximal, for edge-on galaxies) dimming in the 2-10 keV band varies between 20\% (40\%) in the stochastic and remnant phase, to 40\% (55\%) during the merger if the source were at $z=0$, and it is almost negligible for $z=3$. We discuss this further in section 2.2.

An additional  difference between theory and observations  is related to the fact that observed samples do not contain information about the fraction of sources in the stochastic, merger and remnant stages, and how many represent a merger with mass ratio of 1:2, 1:6 or 1:10.  As shown in Fig.~\ref{fig:sfr_bhar_2w_app}, the three stages cover differently the SFR-BHAR plane and a proper comparison with the observations requires an adjustment of the various numbers in the three groups.

The approach we adopt is to specify the boundaries of a region in the BHAR-SFR plane relevant to a population of relatively low-mass galaxies that includes quiescent galaxies and mergers with mass ratio from 1:1 to 1:10. Real correlations depend on the distribution of points within these boundaries, and will include galaxies of different masses and structural parameters, and this we cannot do quantitatively. One can, however, estimate the fraction of objects in each of the merger phases that is included in most observed samples,  e.g.,  the remnant phase in low-redshift samples, and remnants+mergers in higher-redshift samples. The duration of each phase must also be taken into account, since this determines the number of objects from this phase in a specific sample and hence the nature of the derived correlations that depend on such numbers. For example, the merger phase, being short-lived, $\sim 0.2-0.3$ Gyr, is sub-dominant in the general distribution, with respect to the stochastic and remnant phases, which last $0.8-1.3$ Gyr each. 

\subsection{Bi-variate distribution}\label{sec:bivariate}

In Fig.~\ref{fig:sfr_bhar_2w_app} we show the bi-variate distribution of BHAR and SFR within 100~pc, and within 5~kpc, distinguishing galaxies in the stochastic (red), merger (gold) and remnant (dark gray) phases.  In the case of the central SFR, $\sfrn$ (left panel), the stochastic  and remnant phases occupy similar regions, while the merger phase, beyond a primary peak coincident with the other two phases, presents a secondary peak at higher BHAR and $\sfrn$.  The secondary peak is limited in its extent in SFR because averaging over 100~Myr removes the highest peaks of SFR, which are associated to the highest peaks in BHAR (cf. Fig.~14 in  Volonteri et al. 2015). As noted previously  \citep{2010MNRAS.407.1529H,2014MNRAS.443.1125T,2012ApJ...746..168D,2013ApJ...765L..33L,2014ApJ...780...86E} $\sfrn$ is a better tracer of BHAR than $\sfrg$, i.e., they are better correlated.  
As discussed by Capelo et al. 2015 \citep[see also][]{2010MNRAS.407.1529H}, the main drivers of the BHAR are the gas content  \citep[setting an overall `normalisation',  see also][]{2013ApJ...771...63R,2014MNRAS.441.1059V} and local losses of angular momentum, and these became more clearly linked to SFR in the nuclear region. 
We have compared the results of our standard run to a low-resolution run (see Appendix of Volonteri et al. 2015). The degree of correlation is degraded at lower resolution, showing that our simulations are accurately representing inflows due to angular momentum loss. Additionally, because the nuclear region contains both cold inflowing gas and gas affected by thermal feedback, we calculate the accretion for each individual particle separately, rather than averaging over the gas broadly.  The hot gas contribution is fairly negligible, because of its lower density and higher temperature.

The link between galaxy-wide SFR ($\sfrg$, right) and BHAR seems much weaker.  Galaxies in the stochastic phase occupy a region in the BHAR-$\sfrg$ plane (red contours in Fig.~\ref{fig:sfr_bhar_2w_app}) roughly tracing the region between the relations suggested by Mullaney et al. 2012 and Chen et al. 2013 to characterise star-forming galaxies (dotted blue lines). The merging systems (gold contours) approach the upper part of the curves shown in \cite{2012A&A...545A..45R},  who suggest that the highest luminosity AGN are merger-driven, and they exhibit a tighter correlation between BHAR and  $\sfrg$. In fact, in the merger phase,  losses of angular momentum are driven by global dynamics and BHAR and SFR become somewhat better correlated even on large scales.  In the remnant phase (gray contours) a wide range of BHAR can be associated to a given $\sfrg$. 

The remnant phase is, perhaps, the most interesting when comparing to AGN observations,  as 80 percent of the AGN do not show  any hint of a companion \citep[e.g.,][]{2012A&A...545A..45R} and are ordinary (massive) star forming galaxies. For most of the remnant phase the simulated galaxies do not show strong morphological disturbances, however, the BHAR is sufficiently high at times that the galaxies enter the AGN dominated region of the  BHAR-$\sfrg$ diagram.  Observationally, this has been a matter of some debate. Rosario et al. (2013)  find a strong connection between AGN activity and SFR, i.e. most AGN hosts are on the main-sequence of star forming galaxies \citep{elbaz2007,noeske2007}. It has been suggested that there is an excess of AGN in post-merger \citep{2013MNRAS.435.3627E} or post-starburst galaxies \citep{2010MNRAS.405..933W}, and that galaxies hosting moderate-luminosity AGN are transitioning from SF to quiescence (Schawinski et al. 2009).   Our broad interpretation is that the AGN is more likely to be observed in the (much longer) remnant phase, where it enters the AGN-dominated region. In fact, in our calculations (Fig 13 in Volonteri et al 2015), the AGN zigzags continuously in and out of this region.   

We can now estimate the fraction of galaxies in different stages of the merger that are likely to be found in large observed samples. Galaxies in the merger phase would be considered AGN-dominated ($L_{\rm AGN}>L_{\rm SFR}$) for about 40 per cent of that phase, galaxies in  the remnant phase are AGN-dominated for 25 per cent  of their phase, and galaxies in the stochastic phase 7 per cent. However, the merger phase is much shorter. Therefore the overall probability, defined as the ratio between the time when $L_{\rm AGN}>L_{\rm SFR}$ and the total simulation time (stochastic + merger + remnant), of finding a galaxy in the remnant phase in the AGN-dominated region is almost twice as large than for a merging galaxy (13 per cent versus 7 per cent respectively), and six times higher than for a galaxy in the stochastic phase. 
 
 \begin{figure}
\centering
\includegraphics[width=\columnwidth,angle=0]{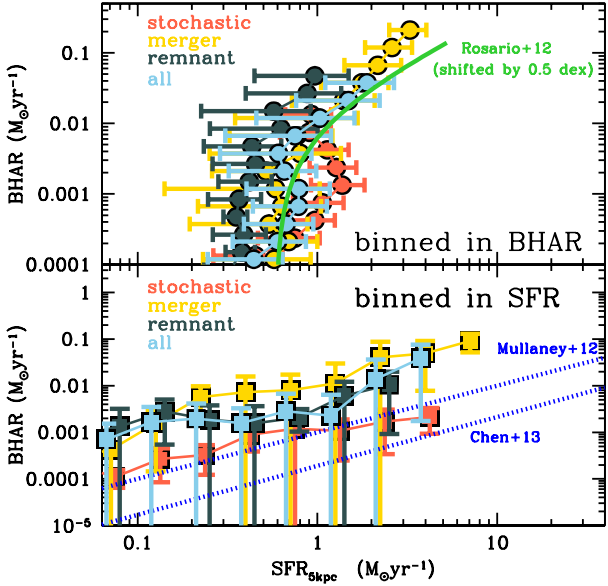}
\caption{ Mean SFR in bins of BHAR (top) and mean BHAR in bins of SFR (bottom); merger and remnant points have been slightly shifted to enable appreciating the size of the one-$\sigma$ error bar. For the comparison, note that Rosario et al. (2012; green) measured the mean SFR for AGN in different luminosity bins, while Mullaney et al. (2012; blue)  Chen et al. (2013; blue) measured the mean BHAR by stacking galaxies in bins of SFR.}
\vspace{-0.5cm}
\label{fig:averages}
\end{figure}

\subsection{BHAR versus SFR using binned data}\label{sec:proj}
 
Most observational papers cannot consider the full bi-variate distribution of SFR and BHAR. Instead, they focus on the  mean SFR for AGN  in different luminosity bins (i.e. stacking  in bins of BHAR; e.g., Rosario et al. 2012) or the mean BHAR in SFR bins (i.e. stacking  in bins of SFR; e.g., Mullaney et al. 2012, Chen et al. 2013 and Delvecchio et al. 2015).
To compare with those studies, we binned our simulations in the same way as used by them:  i.e., we calculate the mean SFR in bins of BHAR and the  mean BHAR in bins of SFR.  Our points are not  uniform on the SFR and BHAR axes across the different evolutionary stages, while observational samples use relatively well distributed bins across their range.  We therefore discard bins with few points ($<50$ for sub-samples, $<300$ for the complete sample), and draw only 50 or 300 random points respectively for the remaining bins.

The results are shown in Fig 2 where the first way of binning is shown in the top panel and the second in the bottom panel.  Remarkably, in the former case we recover the trends of Rosario et al. (2012), while in the latter we recover the trend found by Mullaney et al. (2012),  Chen et al. (2013) and Delvecchio et al. (2015). We suggest that the different trends found for AGN and SF galaxies are in part caused by the different  projections of the full bi-variate distribution,  and that the intrinsic distribution of properties in those samples is similar to the one shown in Fig.~\ref{fig:sfr_bhar_2w_app}. 

In light blue we show a bona fide global sample which takes into account the time spent in each stage, and where  we have enforced high statistical significance in each bin. This is the sample to be compared to observations (SFR $>0.5$~M$_\odot$~yr$^{-1}$ seems also a plausible lower limit for the SFR). The process of averaging, and the statistical significance, have an obvious effect in removing the highest BHAR/SFR sources. For instance, when we require a large number of points per bin, the less populated high BHAR (SFR) bins disappear. Also, the mean in a BHAR (SFR) bin is driven by the more numerous sources at low SFR (BHAR). 

In the top panel, except for the two bins at the highest BHAR,  the SFR is consistent with being uncorrelated with BHAR. The  wiggles show that the weight is dominated by the stochastic phase at low BHAR, then the remnant, and finally the merger phase take over. The bend at high BHAR is driven only by the last two points, dominated by the merger stage, as proposed by Rosario et al. 2012. However, the increase in SFR is only a factor 2 to 3, similar to the enhancement seen by \cite{2015arXiv150207756S}. An increase in the average masses of hosts of the most luminous AGN, or an enforced correlation between SFR and the long term BHAR may also be responsible for the trend.

Note, again, that our simulations apply only to small-to-medium stellar mass systems. We can try to extrapolate the behaviour to more massive galaxies using the arguments outlined in Volonteri et al. 2015, that are based on the assumption that both BHAR and SFR would increase approximately linearly with stellar mass. In this case, larger galaxies would move by about one order of magnitude of BHAR per one order of magnitude of SFR in the plane shown in Fig.~\ref{fig:sfr_bhar_2w_app}.  This  leads us to suggest that if our stellar masses and BHs were two orders of magnitude larger (a few times $\sim10^{12}$ and $10^{8}$ solar masses respectively) then they would occupy a region surrounding the $z=2$ curve of \cite{2012A&A...545A..45R} when we bin in BHAR. In Fig.~\ref{fig:averages} we show the $z=0$ curve shifted
by 0.5 dex in both BHAR and SFR  (the shapes of all these curves in Rosario et al. 2012 are basically identical).  

\begin{figure}
\centering
\includegraphics[width=\columnwidth,angle=0]{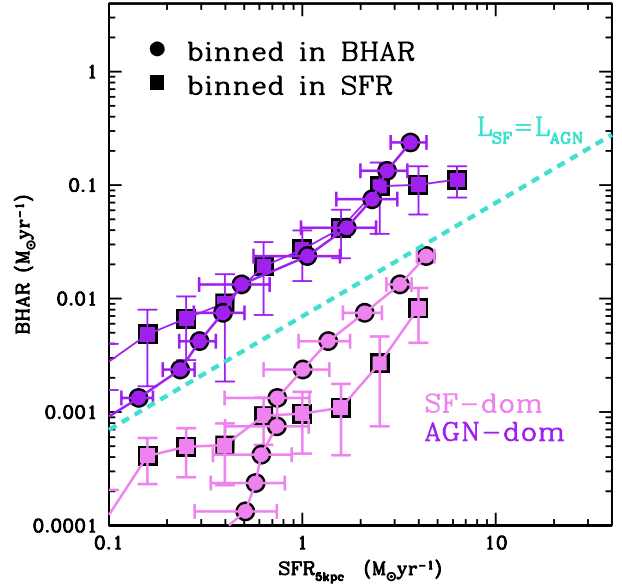}
\caption{Trends of SFR and BHAR for binned data, dividing the sample in AGN-dominated (darker points) and SF-dominated (lighter points). The BHAR at a given SFR  for AGN-dominated galaxies is about 20 times higher than for SF-dominated ones.}
\vspace{-0.5cm}
\label{fig:averages2}
\end{figure}

 When binning in SFR (lower panel of Fig.~\ref{fig:averages}),  galaxies in the merger and remnant phases are $\sim 0.5-1$ dex above the extrapolation of the relation suggested by Mullaney et al. (2012) and Chen et al (2013). The stochastic phase is closer to their fit in normalization, but has a shallower slope, caused by a drop in BHAR at the highest SFR, when, during a SF burst, SN feedback becomes strong enough to affect the gas properties. The full population has a best-fit $\log(\rm{BHAR})=(1.05\pm0.29) \log(\rm{SFR})-2.27\pm0.12$.  If we included obscuration due to gas in the galaxy, the normalization decreases to $-2.43$.  Furthermore, SF-dominated galaxies ($L_{SF} > L_{AGN}$, light squares in Fig.~\ref{fig:averages2}) have $\log(\rm{BHAR})=(0.93\pm0.17) \log(\rm{SFR})-2.89\pm0.07$, slightly shallower and highly normalized than, but very close to, the curve proposed by Mullaney et al. (2012).  Obscuration in the host would further decrease the normalization to $-2.96$. 

 Beyond binning, selection also matters. As highlighted by \cite{2015arXiv150207756S}, their study and \cite{2012A&A...545A..45R} start with a population of AGN. Mass- or SF-selected samples start from a parent population of galaxies, which may or not host an AGN, and the average is likely dominated by sources which do not have AGN. Starting from an AGN population one studies the direct link between BHAR and SFR for galaxies in an active phase. A galaxy parent population instead highlights the broad BHAR-SFR co-evolution over cosmic time and the whole population. In Fig.~\ref{fig:averages2} we propose a new way of looking at the SFR-BHAR connection. Observations often have only upper limits for either SFR or BHAR.  In simulations, instead, we know when the BH is accreting or not, and what the SFR is, down to very low values (e.g., when SF is suppressed by SN feedback after a burst).  We divide our sample in AGN- and SF-dominated galaxies ($L_{SF}$ larger or smaller than $ L_{AGN}$ respectively) and estimate the  increase in BHAR versus SFR in galaxies during the phase where BHs are actively growing. AGN-dominated galaxies have a BHAR-SFR relation with a normalization about 20 times higher than SF-dominated ones, i.e. their BHs grow relatively more than the galaxy stellar mass by this factor. As the BHAR, i.e., the AGN luminosity limit, is lowered, more of the SF-dominated region is sampled, and at low luminosity AGN-selected samples cross from AGN-dominated to SF-dominated, thus lessening, or erasing, the significance of a correlation between SFR and BHAR (see also Hickox et al. 2014).

\section{Conclusions}\label{sec:Conclusions}

We analyse a suite of high-resolution galaxy merger simulations to study  BH and galaxy properties during various phases of the mergers.  We calculate the BHAR and the SFR  during the `stochastic' phase (galaxies in isolation or in the early  phases of an encounter),  the `merger' proper (when the merger dynamics dominates),  and the `remnant' phase (from the end of the merger to the return to quiescence). We find that in the remnant phase the BHAR can be sufficiently high at times to move the galaxy into the AGN dominated region. The  probability of finding a galaxy in the remnant phase, even long after the starburst took place, in the AGN-dominated region is twice as large than for a merging galaxy. The main goal in this Letter is to compare the relationships between BHAR and SFR with observations. 

 We find that different projections of the bi-variate distribution recover different  trends of the population. If the observations can be extrapolated to the simulated mass range, we are able to reconcile seemingly contradictory observational results when the mean SFR for AGN in different luminosity bins (e.g., Rosario et al. 2012) or the mean BHAR stacking galaxies in bins of  SFR are measured (e.g., Mullaney et al. 2012,  Chen et al. 2013, Delvecchio et al. 2015). Hickox et al. (2014), with  an ansatz that over long timescales BHAR and SFR are perfectly correlated, reach similar conclusions, also suggesting that future progress requires a direct measurement of the bivariate SFR/BHAR distribution. 

The bi-variate distribution derived from our simulations suggests that galaxies in the stochastic phase would not be considered AGN-dominated ($L_{SF} \gg L_{AGN}$).   Galaxies would be  AGN-dominated chiefly during the merger and remnant phases.  The BHAR at a given SFR  for AGN-dominated galaxies is about 20 times higher than for SF-dominated ones.  During most of the merger proper and in the remnant phase a given SFR can be associated with a large range of BHAR. A possible exception is the groups of galaxies with the highest SFR and the highest  BHAR, characterised by having $L_{SF}\simeq L_{AGN}$.

\section*{Acknowledgements}
MV thanks E. Daddi, D. Alexander and V. Wild for valuable suggestions and discussions, and G. Mamon for help with supermongo. MV acknowledges  support from NASA (ATP NNX10AC84G), from SAO (TM1-12007X), from NSF (AST 1107675), and a Marie Curie FP7-Reintegration-Grant  (PCIG10-GA-2011-303609). HN acknowledges support by the Israel Science Foundation grant 284/13. 
This work was granted access to the HPC resources of TGCC under the allocations 2013-t2013046955 and 2014-x2014046955 made by GENCI. This research was supported in part by the National Science Foundation under grant no. NSF PHY11-25915, through the Kavli Institute for Theoretical Physics. PRC thanks the Institut d'Astrophysique de Paris for hosting him during his visits.

\scalefont{0.94}
\setlength{\bibhang}{1.6em} 
\setlength\labelwidth{0.0em}
\bibliographystyle{mn2e}

\normalsize

\end{document}